 \newlength\smallfigwidth
\documentclass[aip,jap,reprint,floatfix,showpacs,amsmath,amssymb]{revtex4-1}

\usepackage{graphicx}
\usepackage[hypertex]{hyperref}
\def\ba{\begin{eqnarray}}
\def\ea{\end{eqnarray}}
\def\be{\begin{eqnarray}}
\def\ee{\end{eqnarray}}

\begin{document}

\preprint{UFV}

\title{Curvature-induced changes in the magnetic energy of vortices and skyrmions in paraboloidal nanoparticles}

\author{V. L. Carvalho-Santos}
\affiliation{Departamento de F\'isica, Universidad de Santiago de Chile and CEDENNA,\\ Avda. Ecuador 3493, Santiago, Chile}
\affiliation{Instituto Federal de Educa\c c\~ao, Ci\^encia e Tecnologia Baiano - Campus 
Senhor do Bonfim, \\Km 04 Estrada da Igara, 48970-000 Senhor do Bonfim, Bahia, Brazil}

\author{R. G. Elias}
\affiliation{Departamento de F\'isica, Universidad de Santiago de Chile and CEDENNA,\\ Avda. Ecuador 3493, Santiago, Chile}

\author{J. M. Fonseca}
 \affiliation{Departamento. de F\'isica, Universidade Federal de Vi\c cosa, \\Av. Peter Henry Rolfs s/n, 36570-000, Vi\c cosa, Brazil}

\author{D. Altbir}
\affiliation{Departamento de F\'isica, Universidad de Santiago de Chile and CEDENNA,\\ Avda. Ecuador 3493, Santiago, Chile}

\begin{abstract}
Curvature effects are  important for a proper description of  the properties of magnetic systems. In this paper the exchange and dipolar energy of vortices on a paraboloidal shell is studied. Using analytical calculations it is shown that the in-plane component of vortices has larger energy on a paraboloidal shell than in a planar disk with same thickness. On the other hand, the dipolar energy associated to the vortex core diminishes if the vortex core is on a paraboloidal surface. This reduction in the dipolar energy may cause a vortex pinning mechanism by a paraboloidal shaped defect in a planar nanomagnet. Regarding skyrmions, by using an in-plane anisotropy approximation to the dipolar energy, it is shown that the skyrmion must have its width shrunk in order to diminishes the magnetostatic energy and satisfy geometrical constraints of the system. 
\end{abstract}
\maketitle

%
%


\section{Introduction}
Magnetic vortices consist of a spin closure texture that can appear as the magnetization groundstate of circular nanomagnets such as disks \cite{Guslienko-JAP-2004,Usov-JMMM-1993}, rings \cite{Klaui-Review,Castillo-APL-2014,Vaz-PRB-2006}, spheres \cite{Kravchuk-PRB-2010} and torus \cite{Vagson-JAP-2010}. Studying and controlling the stability, polarity and chirality of vortices as well as their pinning mechanisms are important issues, once these structures have been considered as candidates to compose data storage and magnetic access memory devices \cite{Hertel}. In order to understand the dynamical properties of vortices, it has been shown that normal mode magnon frequencies for vortices in magnetic dots differs from the case of uniformly magnetized dots \cite{Ivanov-APL-2002} and a doublet splitting of the high frequency modes, which can be amplified by an external field, can occur for both dots and rings \cite{Ivanov-PRB-2005}. Furthermore, the study of the influence of dipolar magnetic fields on the spin mode frequencies of dots showed that the local dipolar approximation leads to an underestimation of the frequencies \cite{Zivieri-PRB-2008}.

From another side, curvature plays an important role to describe the properties of nanomagnetic particles. In this contest, a lot of effort has been done in order to study geometry effects on the properties of nanomagnets. For instance, unlike what happens for planar circular nanodots, the critical magnetic field which induces a vortex core switching in magnetic hemispherical caps depends on the vortex chirality \cite{Sloika-APL-2014}. More recently, a magnetic energy functional for an arbitrary curved thin shell has been developed and the authors showed that curvature acts as an effective magnetic field into the system \cite{Gaididei-PRL-2014}. From experimental point of view, curved nanostructures such as permalloy caps on non-magnetic spheres \cite{Streubel-APL-2012}, cylindrically curved permalloy magnetic segments with different radii of curvature on non-magnetic rolled-up membranes \cite{Streubel-Nanoletters-2012} and periodically modulated nanowires \cite{knielsch-works} have been produced. Although paraboloidal nanomagnets have not yet been developed, it was shown that computer assisted design models can be used to construct ellipsoidal shapes \cite{Lazarus-PRL-2012}. Thus, by using atomic layer deposition techniques, we can expect that paraboloidally-shaped magnetic structures can soon be produced.

Regarding theoretical works, excitations coming from the Heisenberg model on a paraboloidal surface have been previously studied \cite{Boas-PLA-2014}. It was shown that the in-plane vortex component presents larger energy on the paraboloid surface in relation to the planar case. However, in order to avoid divergences in the magnetic energy, magnetic vortices develop an out-of-plane component, so called vortex core, whose radius predicted by micromagnetic simulations has a diameter in the order of $50$ nm \cite{Guslienko-JAP-2004}. The development of the vortex core yields a magnetostatic energy cost, coming only from surface and volumetric magnetic charges, formally defined as $\sigma=\mathbf{m}\cdot\hat{n}$ and $\varrho=\nabla\cdot\mathbf{m}$, respectively. Thus, in order to analyze the stability and vortices pinning and depinning mechanisms induced by curved defects, the dipolar energy associated of the vortex core must be taken into account. 

Based on these ideas and assuming that the geometric parameters of the paraboloid allow the appearance of a vortex as the magnetization groundstate, we extend the results of Ref. [13] in its analysis about the skyrmions characteristic length on a paraboloidal surface by calculating the anisotropy-like dipolar cost to maintain a skyrmion on a paraboloidal shell.

\section{ Model}

For our purposes, it will be convenient to parametrize a magnetic paraboloid with thickness $L$ using a parabolic coordinates system
\begin{eqnarray}\label{Parametrization}
\mathbf{x}=\xi\eta\cos\varphi\,\mathbf{x}+\xi\eta\sin\varphi\,\mathbf{y}+\frac{1}{2}(\xi^2-\eta^2)\,\mathbf{z}\,,
\end{eqnarray} 
where $\mathbf{x}$, $\mathbf{y}$ and $\mathbf{z}$ are the unitary vectors in Cartesian coordinates. In order to get a paraboloid of revolution opening downward, we have $\varphi\in[0,2\pi]$, $\eta\in[-\eta_{_R},\eta_{_R}]$ and $\xi\in[\xi_0,\xi_0+\delta\xi]$, with $\xi_0>0$ being the value of the coordinate $\xi$ when $\eta=0$ and $\delta\xi=\sqrt{2L+\xi_0^2}-\xi_0$, with $L$ being the thickness of the paraboloidan nanoparticle. The metric elements are $g_{\varphi\eta}=g_{\eta\varphi}=0$, $g_{\varphi\varphi}=\xi^2\eta^2$ and $g_{\eta\eta}=g_{\xi\xi}=\xi^2+\eta^2$. The surface and the volume elements are given by $dS=\xi\eta\sqrt{\xi^2+\eta^2}d\varphi d\eta$ and $dV=\xi\eta(\xi^2+\eta^2)d\varphi d\xi d\eta$, respectively and the unitary normal vector is $\mathbf{n}=(\eta\hat{\rho}+\xi\hat{z})/\sqrt{\xi^2+\eta^2}$, where $\hat{\rho}=\hat{x}\cos\varphi+\hat{y}\sin\varphi$. This parametrization allows us to continuously deform the paraboloid in a plane by taking $\eta/\xi\rightarrow0$, where we have defined the paraboloid radius as $r\equiv\xi\eta$.

In the absence of anisotropy and Zeeman interactions, the magnetic energy is $E=E_{\text{ex}}+E_{\text{dip}}$, where $E_{\text{ex}}$ is the exchange energy, given by
\begin{eqnarray}\label{heiscont} 
E_{\text{ex}}=A\int g^{\mu\nu}{\partial_\mu m^{\alpha}}{\partial_\nu 
m_{\alpha}} dV\,,
\end{eqnarray} 
where $A$ is the stiffness constant and
\begin{eqnarray}\label{dipolar}
E_{\text{dip}}=-\frac{\mu_0M_{S}^2}{2}\int_{V}m^\alpha\,\mathbf{x}_\alpha\cdot \mathbf{e}_\mu\,g^{\mu\nu}\partial_\nu\Phi\,dV\,,
\end{eqnarray}
is the magnetostatic energy, with $\mu_0$ being the magnetic permeability. In the above set of equations $\Phi$ is the magnetostatic potential and the magnetization vector is parametrized in Cartesian coordinate system, $\mathbf{m}=\mathbf{M}/M_{_S}\equiv(m_1,m_2,m_3)$, where $M_{_S}$ is the saturation magnetization. The surface is described by the curvilinear coordinates $\eta$, $\xi$ and $\varphi$. $g^{\mu\nu}$ is the surface space contravariant metric, $\mathbf{x}_\alpha$ and $\mathbf{e}_\mu$ describes the unitary vectors in Cartesian and in the curvilinear basis, respectively. In eq (\ref{heiscont}) and (\ref{dipolar}) the Einstein summation convention is adopted, with $\mu$, $\nu$ and $\alpha$ varying from 1 to 3.

The vortex configuration will be described by a magnetization that is independent of $z$ and $\varphi$, that is, 
\begin{eqnarray}
\mathbf{m}=m_{z}(r)\hat{z}+m_{\varphi}(r)\hat{\varphi}\,,
\end{eqnarray}
where the function $m_z(r)\equiv m_z$ specifies the core profile. Here we will consider a rigid vortex core model \cite{New-Ref-Dora}, in which there are no changes in the size and form of the core profile due to the curvature. To describe the core profile, we adopt the model proposed by Landeros \textit{et al} \cite{Landeros-PRB-2005}, in which 
\begin{eqnarray}
m_{z}=\left\{\begin{array}{ll}\left[1-\left({r}/{r{_c}}\right)^2\right]^n\,, & r<r_c\nonumber\\
0\,\,, & r>r_c,\end{array}\right.\,
\end{eqnarray}
where $r{_c}\equiv\xi\eta_c\leq R$ is the vortex core radius, $R\equiv\xi\eta_{_R}$ is the maximum radius of the paraboloid and $n$ is a non-negative integer. The azimuthal component can be determined from the normalization condition $m_\phi^2+m_z^2=1$. It can be observed that despite the proposed magnetization configuration varies along the volume of the nanomagnet, we have that $\varrho=0$ and so, only the surface magnetic charges accounts to the magnetostatic energy calculations.
 
\begin{figure}
\includegraphics[scale=0.20]{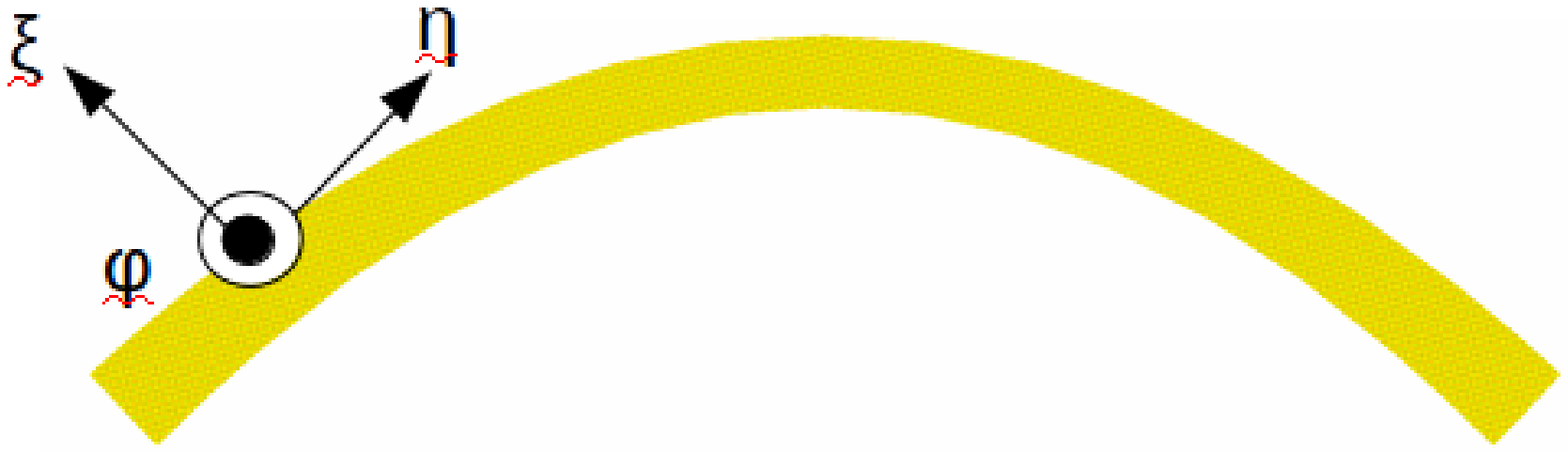}\includegraphics[scale=0.30]{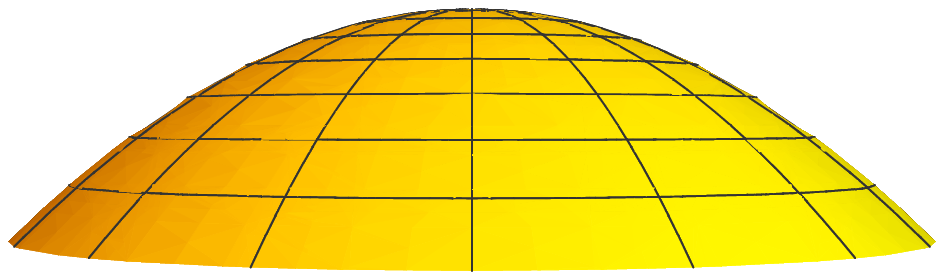}\\
\includegraphics[scale=0.30]{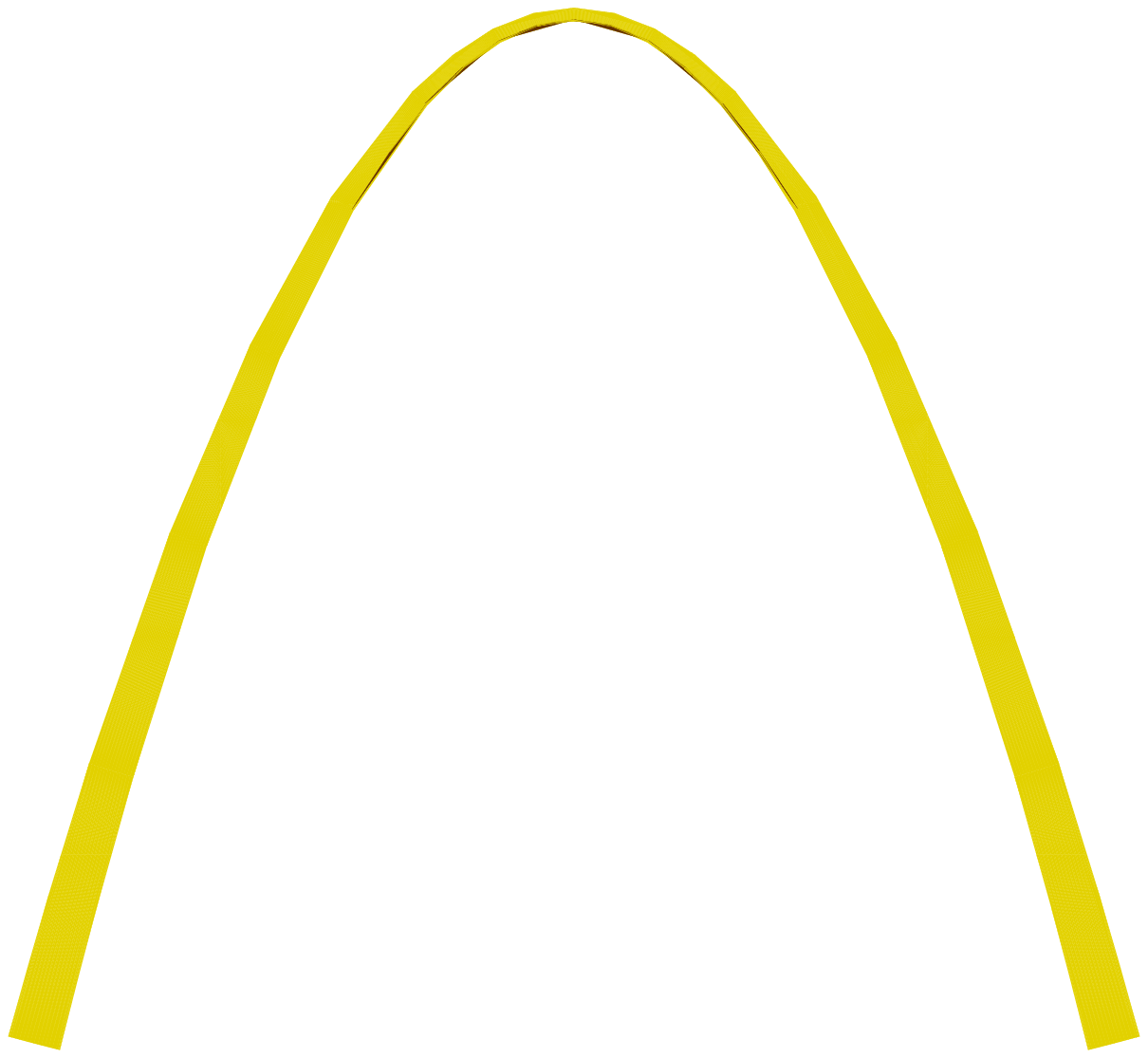}\includegraphics[scale=0.30]{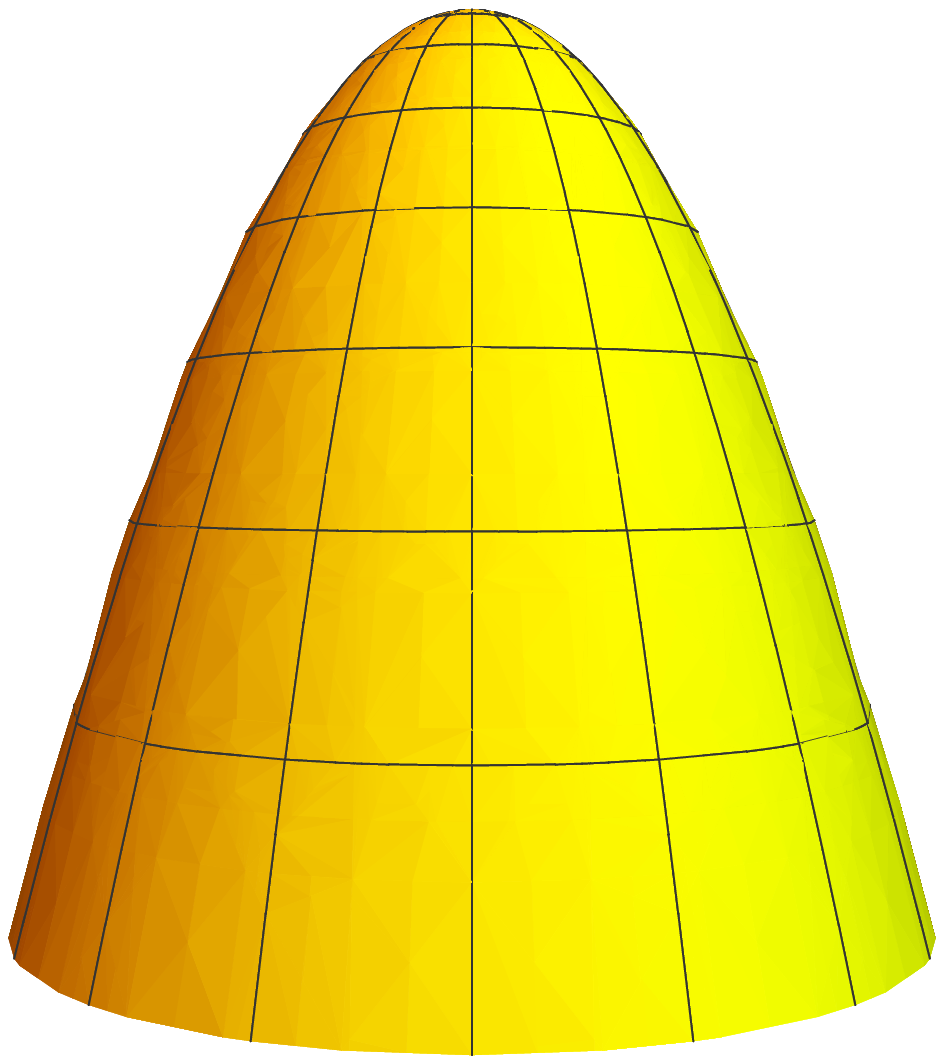}
\caption{[Color online] Paraboloidal shell described by Eq. (\ref{Parametrization}). Upper figures show a transversal cut and a paraboloid with $\xi_0/\eta_{_R}=1$. Lower figures show a transversal cut and a paraboloid with $\xi_0/\eta_{_R}=1/4$.}\label{ParabShape}
\end{figure}

\section{Results}
By applying the described model on the paraboloidal surface, the exchange energy associated with the vortex configuration is
\begin{eqnarray}\label{VortexPatternEnergy}
\frac{E_{_\text{ex}}}{2\pi A}=E_{_C}
+(\eta_{_R}^2-\eta_c^2)\ln\left(\frac{\sqrt{2L+\xi_0^2}}{\xi_0}\right)\nonumber\\+L\ln\left(\frac{R}{r_c}\right)\,,
\end{eqnarray}
where $E_{_C}$ is the vortex core exchange energy. Since we are leading with a rigid core model, the vortex core exchange energy does not depend on the geometry, being evaluated as   \cite{Landeros-PRB-2005}
\begin{eqnarray}
E_{_C}=L\left(\frac{1}{2}H[2n]-nH[-\frac{1}{2n}]\right)\,,
\end{eqnarray}
where 
\begin{eqnarray}
H[z]=\sum_{i=1}^{\infty}\left(\frac{1}{i}-\frac{1}{i+z}\right)\,
\end{eqnarray}
are the harmonic numbers \cite{Landeros-PRB-2005}.

The third term in the right of Eq. (\ref{VortexPatternEnergy}) is the exchange energy associated to the in-plane component of the vortex on a plane disk with height $L$. Then, the in-plane component of the vortex energy depends on the geometrical parameters of the paraboloid such that the greater the relation $\eta_{_R}/\xi_0$, the larger the vortex energy. Notably, lowest vortex exchange energy is obtained for $\eta_{_R}/\xi_0\rightarrow0$ in such a way that the second term in the right of Eq. (\ref{VortexPatternEnergy}) vanishes. Thus, the vortex exchange energy is always larger on the paraboloid than the energy in its cylindrical counterpart. 

\begin{figure}
\includegraphics[scale=0.25]{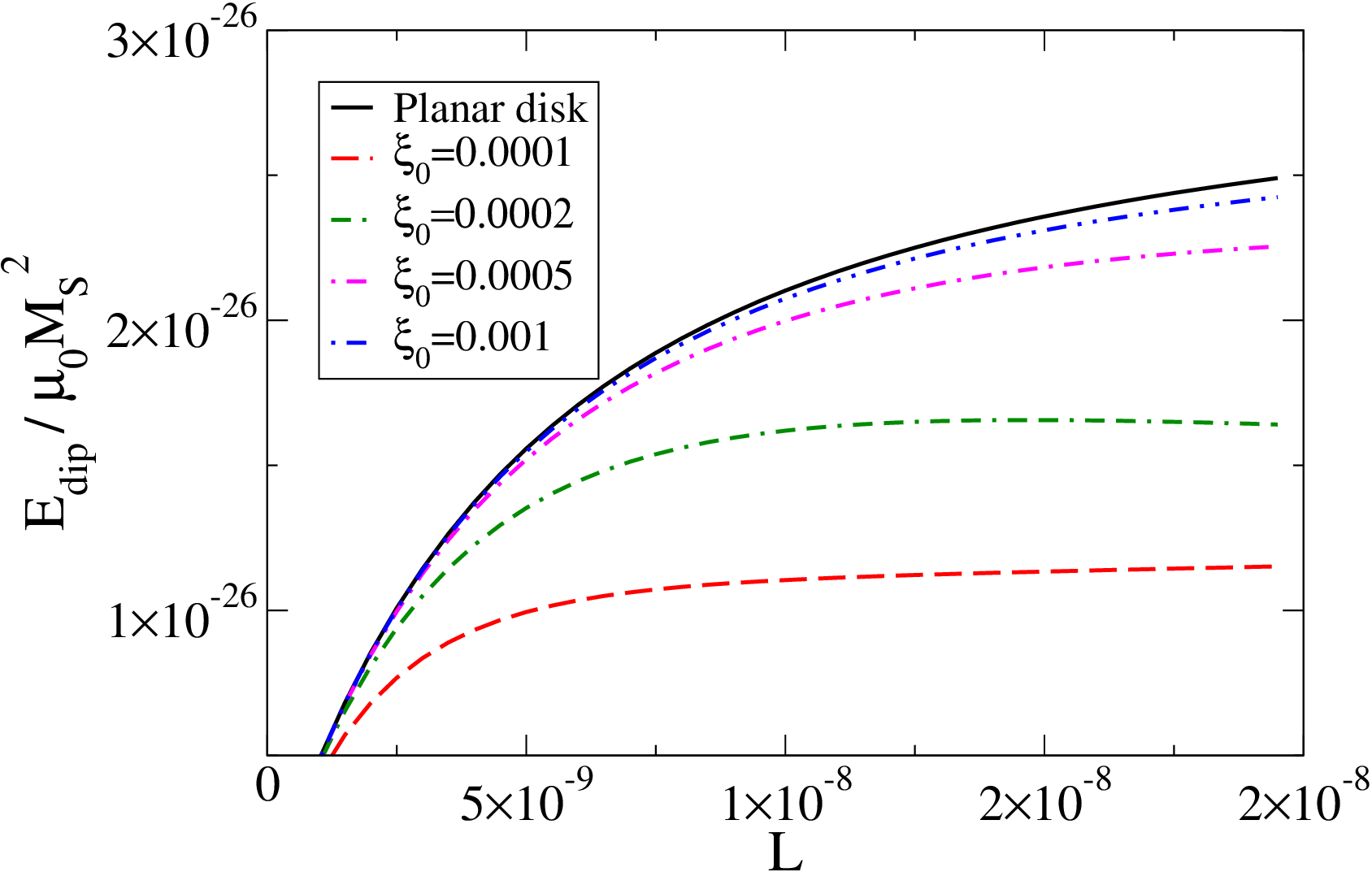}\caption{Normalized dipolar energy of the vortex core in the paraboloid in function of $L$ for different values of $\xi_0$. }\label{EnergyGraphycs}
\end{figure}

Aiming calculate the dipolar energy of the vortex core on the paraboloid, we must determinate the magnetostatic potential, which can be calculated from expanding the inverse of the distance in parabolic coordinates \cite{Goo-RMF-1998}. Defining $m_z\,'\equiv m_z(r\,')$, after some manipulation, the magnetostatic potential is evaluated as (See Supplemental Material \cite{Supp-Mat} for more details)
\begin{eqnarray}
\Phi=\int_0^{\eta_c} {m_z\,'\eta\,'}\,d\eta\,'\int_0^\infty dk\, k\,\Lambda(\xi)\,J_0(k\eta\,')J_0(k\eta)\,,
\end{eqnarray}
where $\Lambda(\xi)=\xi_1^2I_0(k\xi)K_0(k\xi_1)-\xi_0^2 I_0(k\xi_0)K_0(k\xi)$, $J_0$ is the Bessel function of first kind, $I_0$ and $K_0$ are the modified Bessel functions of first and second kind, respectively. Taking this result into Eq. (\ref{dipolar}), we find
\begin{widetext}
\begin{eqnarray}\label{DipEnergy}
\frac{E_{\text{dip}}}{\mu_0M_{S}^2}=\pi\int_0^\infty dk\,k\left(\frac{2}{k}\right)^n\frac{\eta_c^{-n+1}}{k}J_{n+1}(k\eta_c)\Gamma(n+1)\Biggl\{\left(\frac{2}{k}\right)^n\frac{\eta_c^{-n+1}}{k}J_{n+1}(k\eta_c)\Gamma(n+1)\nonumber\\\times\biggl[\Psi(k)-\frac{2}{k}\,\Xi(k)\biggl]
-{2^{n}\eta_c^3}(k\eta_c)^{-(n+1)}J_{n+2}(k\eta_c)\Gamma(n+1)\,\Xi(k)\Biggl\}\,\,\,\,\,\,\,\,
\end{eqnarray}
\end{widetext}
where 
\begin{eqnarray}
\Psi(k)=\xi_1^4I_0(k\xi_1)K_0(k\xi_1)+\xi_0^4I_0(k\xi_0)K_0(k\xi_0)\nonumber\\
-2\xi_1^2\xi_0^2I_0(k\xi_0)K_0(k\xi_1)\,\,\,\,\,\,\,\,
\end{eqnarray}
and 
\begin{eqnarray}
\Xi(k)=\xi_1^3I_1(k\xi_1)K_0(k\xi_1)-\xi_1^2\xi_0I_1(k\xi_0)K_0(k\xi_1)\nonumber\\+
\xi_0^2\xi_1I_0(k\xi_0)K_1(k\xi_1)-\xi_0^3I_0(k\xi_0)K_1(k\xi_0)\,\,\,\,\,\,\,\,
\end{eqnarray}
The remaining integral can be transformed into a series of hypergeometric functions of various variables, however the obtained expression can not be evaluated with our present computational numeric tools. Due to space limitations, the methodology used to obtain these results is presented in the supplemental material accompanying this paper \cite{Supp-Mat}. Therefore, we solve numerically Eq. (\ref{DipEnergy}). The main results are shown in Fig. \ref{EnergyGraphycs}, that illustrates the behavior of the dipolar energy of the vortex core on the paraboloid in function of the thickness of the particle for some values of $\xi_0$. It can be observed that the dipolar energy increases with $\xi_0$ in such way that when $\xi_0\gtrsim0.003$, the energy approximates to the value of the energy of the vortex core in a planar disk. It can be also observed an increasing in the vortex core energy by rising the thickness of the nanoparticle. This curvature-induced reduction in the dipolar energy of the vortex core is associated to the reduction of surface magnetostatic charges on the paraboloid. We have also analyzed the case in which the paraboloid is obtained from deforming a disk with radius $r_c$, however, qualitative changes are not observed in this case. 

Due to the reduction in the dipolar energy when the nanomagnet is curved, there must be an increasing in the vortex core radius in order to reduce the associated exchange energy. In this way, the new core radius is obtained from the interplay between magnetostatic and exchange energy in such a way that, by minimizing the total energy, one can get the new vortex core radius. This issue is under preparation \cite{Vagson-Under-Investigation}. Another consequence of this result is that  by assuming that curved defects appearing in planar nanomagnets with a vortex state can be modeled by the paraboloidal shape, the reduction of the dipolar energy when a moving vortex core is on a curved region of the planar device may be behind the pinning mechanisms of vortices by curved defects.

Now we analyze the effect of the dipolar energy on the width of a skyrmion on a paraboloidal shell. Although physically realizable curved thin nanostructures have thicknesses $L$, we follow the formalism developed in Ref. [9] and derive an effective two-dimensional exchange energy of a paraboloidal shell as a limiting case of $L\ll l_{\text{ex}}$, where $l_{\text{ex}}$ is the exchange length of the material. In a previous work, it has been shown that skyrmions can appear as excitations of the Heisenberg system on a paraboloidal surface and a geometrical frustration coming from the two characteristic lengths of the paraboloid leads to a well defined width of the skyrmion  \cite{Boas-PLA-2014}. However, the dipolar interaction brings a new characteristic length into the system in such way that the skyrmion must adopt a new width due to the interplay between curvature and a dipolar-induced in-plane anisotropy. Thus, in order to simplify our analysis, we will reduce the magnetostatic energy as an effective in-plane anisotropy given by
\begin{eqnarray}\label{DipAni}
E_{\text{dipani}}=\omega L\iint (\mathbf{m}\cdot\mathbf{n})^2 dA\,,
\end{eqnarray}
where $\omega$ is a positive constant proportional to $\mu_0M_S^2$. It was shown that this is a good approximation to calculate the magnetostatic energy of thin shells with in-plane magnetization configurations \cite{Slastikov-MMMAS-2005}. Despite the lack of a more systematic work showing that this kind of approximation is valid when the magnetization points to the out-of-plane direction, for our purposes it is enough to assume that Eq. \ref{DipAni} represents  the dipolar energy of the skyrmion magnetization configuration. A more detailed report about the effects of the dipolar interaction on the skyrmion in a curved manifold is under preparation \cite{Vagson-Under-Investigation}.

When only the exchange energy is taken into account, a skyrmion can be well represented by Eq. (13) in Ref. [13], in which $m_{z}={1}/{\sqrt{1+\zeta^2}}$, where
\begin{eqnarray}
\zeta=\sqrt{1+\frac{\eta^2}{\xi_0^2}}-\ln\left[\frac{1}{\eta}\left(1+\sqrt{1+\frac{\eta^2}{\xi_0^2}}\right)\right]\,.
\end{eqnarray}
In this context, the skyrmion width is $\lambda=(1+1/\xi_0^2)^{-1}$. However, in the presence of the interaction given by Eq. (\ref{DipAni}), the new skyrmionic profile is evaluated as  
\begin{eqnarray}
m_{z}=\frac{1+\omega}{\sqrt{(1+\omega)^2+{\zeta^2}}}\,,
\end{eqnarray}
and the width of the skyrmion is evaluated as 
\begin{eqnarray}
\lambda=\left(\frac{1}{1+\omega}+\frac{1}{\xi_0^2}\right)^{-1}\,.
\end{eqnarray}
The new characteristic length introduced by the dipolar energy reduces the skyrmion width because the spins try to align along the surface. Thus, due to the interplay between curvature and anisotropic-like dipolar energy, the skyrmion is shrunk to a smaller region of the paraboloid. It is worth to note that the skyrmion width in a paraboloid is smaller than the one in a planar device, showing the possibility of controlling the skyrmion width by curving planar devices.

\section{Conclusions}

We have calculated the magnetostatic and exchange energy of a vortex on a paraboloidal shell. The competition between exchange and dipolar energy plays an important role on the stability of the vortex in curved nanomagnets. While the exchange energy associated with the in-plane configuration of the vortex is larger on the paraboloid than in a planar counterpart, the dipolar energy associated to the vortex core is smaller in the paraboloidal case. In this way, if a paraboloidal defect is present on a planar magnet in which a vortex is moving, the vortex equilibrium on the defect is possible when the total energy is lower than that one associated to a vortex on a plane geometry. Then, since the magnetic energy of the vortex core is larger when it is in a planar region of a nanomagnet, the reduction of the energy induced by the curvature may be associated to pinning and depinning mechanisms of vortices.

Furthermore, we have analyzed the effect of an anisotropic-like dipolar interaction in the width of a skyrmion  on a paraboloidal shell. It was shown that this energy yields  a reduction of the skyrmion width by the introduction of a new characteristic length into the system. Then, the interplay between curvature and dipolar energy can be used to control the skyrmion width, once a smaller skyrmion radius appears in a paraboloid when compared to a planar device. This result opens the possibility of controlling the skyrmion width by  curvature.

\section*{Acknowledgements}
V.L.C.S. thanks the Brazilian agency CNPq (Grant No. 229053/2013-0), for financial support. D.A. acknowledges the support of FONDECYT under projects 1120356, the Milennium Science nucleus ``Basic and Applied Magnetism'' P10-161-F from MINECON, and Financiamento Basal para Centros Cient\'ificos y tecnol\'ogicos de Excelencia, under project FB 0807. We also acknowledge AFOSR Grant. No. FA9550-11-1-0347. R.G.E. thanks Conicyt Pai/Concurso Nacional de Apoyo al Retorno de Investigadores/as desde el Extranjero Folio 821320024. J.M.F. thanks the support of FAPEMIG.





\end{document}